# Unification of the MWI formalism and Bohmian mechanics for the ensembles of event universes in Minkowski-like space

Oded Shor PhD,[1,3] * & Felix Benninger MD[1,2,3 #] & Andrei Khrennikov, PhD[4 #]

[1]Felsenstein Medical Research Centre, Petach Tikva, Israel; [2]Department of Neurology, Rabin Medical Centre, Petach Tikva, Israel; [3] Faculty of Medicine, Tel Aviv University, Tel Aviv, Israel; [4]Faculty of Technology, Department of Mathematics, Linnaeus University, Vaxjö, Sweden

**Corresponding author (*):**

Oded Shor, Felsenstein Medical Research Centre, Petach Tikva, Israel; shor.oded@gmail.com
**# Equally contributing last authors**

**Email of all authors:**

| | |
|---|---|
| Oded Shor | shor.oded@gmail.com |
| Felix Benninger | benninger@tauex.tau.ac.il |
| Andrei Khrennikov | Andrei.khrennikov@lnu.se |

**Abstract** Diversity of interpretations of quantum mechanics is often considered as a sign of foundational crisis. In this note we proceed towards unification the relational quantum mechanics of Rovelli, Bohmian mechanics, and many worlds interpretation on the basis so called *Dendrogramic Holographic Theory* (DHT). DHT is based on the representation of observed events by dendrograms (finite trees) presenting observers subjective image of universe. Dendrograms encode the relational hierarchy between events, in applications they are generated by clustering algorithms; an algorithm with the branching index p >1 generate p-adic trees. The infinite p-adic tree represents the ontic event universe. We consider an ensemble of observers performing observations on each other and representing them by p-adic trees. In such ``observers universe'' we introduce a kind of Minkowski space structure, which is statistical by its nature. This model unites the observer/system discrepancy. Measurements are performed by observers on observers. Such ``observers universe'' is dynamically changing and is background independent since the space itself is emergent. And within this model, we unify the aforementioned interpretations.

**Key Words:** p-adic numbers; dendrograms; dendrogramic holographic theory; Many worlds interpertation; Bohemian mechanic, Minkowski-like space ; event-universe.

1. Introduction

*Dendrogramic Holographic Theory* (DHT) is a unifying theory, developed through a series of studies [1–7], which adopts a relational-event-observational approach to physics and the broader realm of natural science [8,9] . Its foundational postulate, driving all the theoretical consequences, centers around the adherence to Leibniz's Principle.

1.1. P-adic treelike formalization of Leibniz's Principle

Leibniz's Principle [10], or the Principle of the Identity of Indiscernibles, is commonly expressed as follows*:*

*If, for every property F, object x possesses F if and only if object y possesses F, then x is identical to y. In symbolic logic notation, this can be represented as $\forall F(Fx \leftrightarrow Fy) \rightarrow x = y$. In simpler terms, if x and y are distinct entities, there must be at least one property that distinguishes them from each other.*

DHT is a theory concerning the information derived from observations made by different observers on various events. According to Leibniz's Principle, if a particular observer is unable to distinguish any differences between two observations of distinct events based on any feature, then these two events are considered identical to that observer.

This is the good place to remark that, in fact, this is the essence of indistinguishability treatment for quantum particles: if all observational quantities coincide, say the energy, spin, and so on, then particles are indistinguishable. Surprisingly, we have never seen any reference to this principle in quantum foundational discussions on indistinguishability.

When an observer seeks to differentiate between certain observations within a set of events, they must employ a series of inquiries. These inquiries can take the most straightforward yes/no form (although they can involve any number of possible answers). Mathematically, these sequences of questions can be depicted as trees, akin to decision trees, which are referred to as dendrograms. To quantify the relationship between two events, an observer can utilize the initial question in the sequence where the answers diverge.

In the realm of an endless array of events, the depiction of the inherent nature of these events takes the form of an infinite tree. Among the various types of such trees, there exists a class known as p-adic trees, characterized by their homogeneity and the presence of p > 1 edges emanating from each vertex. These trees exhibit an inherent algebraic structure and a corresponding topology that aligns with this structure [11].

The p-adic topology is characterized by the p-adic ultrametric, which satisfies to the strong triangle inequality. As a result, this field possesses a peculiar geometry where all triangles are inherently isosceles. In this context, the p-adic distance between two branches of the tree is determined by their shared root, with a longer shared root (indicating a higher number of questions answered with the same response) corresponding to a shorter distance. Furthermore, by defining "open" and "closed" balls as follows:

$$B_-(R; a) = \{x : r_p(a, x) < R\}, B(R; a) = \{x : r_p(a, x) \leq R\}$$

Where $r_p(a, x) = |a - x|_p$ is the p-adic distance between the points a and x.

Since branches symbolize distinct events, the event space, whether finite or infinite in nature, is furnished with a p-adic ultrametric. This connection between DHT and p-adic analysis intertwines with the realms of theoretical physics [12–27]. This relationship includes investigations by Parisi et al. [26] into complex disordered systems within the p-adic framework and the broader ultrametric context, as expounded in article [27]. General trees, in general, feature an ultrametric topology, and such topological spaces have found extensive utility in the theory of complex disordered systems [16,28].

Embracing the Leibniz principle carries a significant implication: the necessity to differentiate events through the aforementioned question-based process. This process, in a natural sense, finds its representation within the framework of the p-adic number field. From this question-based procedure, which takes shape as a p-adic tree, emerges an immediate association between each event and all others. Consequently, the Leibniz principle leads directly to Machian relationism, not as a mere assumption (in contrast to theories such as shape dynamics and Brans-Dicke theory [29–31]), but as an intrinsic outcome of the p-adic tree representation of events. Furthermore, endorsing the Leibniz principle results in a background-independent theory, akin to

theories like shape dynamics, loop quantum gravity, spin foams, and causal set theory [29,30,32–34]

## 1.2. Coupling to Rovelli's relational quantum mechanics

It's important to highlight that DHT does not fall within the confines of either the quantum or classical paradigms. Both these paradigms naturally emerge from the p-adic relational tree, as demonstrated in [1–3,5], without the need for any additional assumptions other than the acceptance of the Leibniz Principle.

Notably, *Rovelli's Relational Quantum Mechanics* (RQM) shares a significant ideological similarity with DHT [35]. However, RQM posits that all systems are inherently quantum systems. Like DHT, RQM leverages the concept that any quantum mechanical measurement can be deconstructed into a series of yes-no questions, which is then used to formulate the state of a quantum system (relative to a given observer, much like in DHT).

In contrast to DHT, RQM asserts the completeness of quantum mechanics. Accordingly, RQM posits that there are no hidden variables or additional factors that need to be introduced into quantum mechanics, based on current experimental evidence. As demonstrated in [5], quantum theory can be viewed as an emerging theory stemming from a relational structure. Consequently, notions such as completeness and hidden variables become irrelevant. From this perspective, the various interpretations of quantum mechanics can be seen as corresponding to different emergence frameworks for quantum theory from an event-based relational structure. RQM also includes two empirical postulates:

Postulate 1: There exists a maximum amount of relevant information that can be extracted from a quantum system. DHT aligns with this by asserting that the maximal information about an event is encoded within the event branch contained in the infinite p-adic tree.

Postulate 2: It is always possible to obtain new information from a system. In a similar vein, DHT acknowledges that by introducing more events into the relational structure, additional information can be gleaned from each event. Mathematically this process is described as adding and elongating branches of a dendrogramic tree.

## 1.3. Coupling to Smolin's approach to emergence of quantum mechanics

It's worth mentioning Smolin's exploration of the emergence of quantum mechanics and spacetime, which introduces rationalism perspectives, as evident in works such as energetic causal set theory and [36–39]. While these works share a similar ideology with DHT, they are constructed based on concepts like momentum, energy, and even coordinates. Interestingly, these elements are entirely absent in the foundational construction of DHT, yet they can still potentially emerge from the relational p-adic structure. This is the important foundational advantage of DHT: space-time is not considered as the basic concept, but it is derived from treelike structure of events collected by an observer. In this way, spacetime is subjective – observer dependent. [3,6,7]

Additionally, we'd like to highlight certain parallels between our approach, which leads to the emergence of quantum theory from DHT, and the neural network model of the universe as developed in articles [40–44].

**1.4 Observer-dependent subjective knowledge of universe**

Within the framework of DHT, it is possible for two or more observers to hold distinct epistemic perspectives of the universe they are observing, characterized by differing relational structures. Conversely, two or more observers can share identical epistemic views of the universe, even when their observations encompass different events, signifying that these observers possess equivalent inter-relationships among the events they have measured. In this context, it becomes evident that an observer inherently maintains a subjective viewpoint of the universe, regardless of whether the nature of the universe is classical or quantum. This alignment with Bohr's principle of complementarity and event physics is noteworthy. Bohr consistently emphasized that the outcomes of physical observables do not constitute objective properties of systems but are instead generated during the measurement process. Furthermore, this observer-dependent perspective of the universe is in harmony with the principles of special relativity, wherein the measurement of events in spacetime is contingent upon the observer's momentary inertial frame. If we characterize an observer as its world line curve (accelerated or not) on the background of spacetime, each observer will ultimately obtain different ontic relational view of the universe (two observers with same world line are not allowed as they are the same observer if we postulate Leibnitz principle). This view is dependent on the information about events, localized in the background of spacetime, transmitted to the observer moving with acceleration or without it in spacetime

Thus, information on the universe is strictly observer-dependent. In the DHT framework, measurements can be "subjective" in the sense of quantum mechanics and/or observer-dependent in the sense of special relativity, but both lead to a subjective knowledge of the universe.

In recent work [6] we explored the dynamical laws of subjective knowledge acquired by observers on a system, or universe. Our aim was to bring closer the event-observational approach with the principles of spacetime physics of special relativity. For that purpose, we constructed some spacetime model that uniquely defined a certain subjective relational-informational view of the universe. Moreover, this spacetime possessed a casual structure like the light-cone of the Minkowski spacetime. The real parameter spaces discovered in this study while related to an ensemble of observers, primarily represent purely observer-subjective and observer-dependent knowledge of an observer about the universe. In that sense these spaces are inherently subjective.

DHT is more than just an abstract theoretical framework; it has direct applications to experimentation. Experimental data from various sources can be transformed into dendrograms with the assistance of clustering algorithms. The temporal evolution of data collection by an observer is depicted as dynamics within the configuration space of dendrograms, as detailed in articles [1–3,8,9], with a particular focus on the dynamics discussed in article [4]. We remark that, although different clustering algorithms can generate different dendrograms

It's important to emphasize that DHT remains invariant when employing different clustering algorithms. In other words, the fundamental properties of the dynamics are not contingent upon the specific choice of algorithm, although the visualization through dendrograms may vary. For simplicity, we will proceed using 2-adic yes-no clustering algorithms that generate homogeneous 2-adic trees, featuring one incoming branch and two outgoing branches for each vertex.

**1.5. emergence of MWI**

In this paper, we demonstrate the emergence of the Many Worlds Interpretation (MWI) of quantum theory within the framework of DHT. Building on our previous work, which illustrated the emergence of single-observer quantum mechanics from DHT—a correspondence that aligns with the Bohmian interpretation [5]—the emergence of MWI naturally follows when we consider the interactions of many observers as they continually refine their subjective dendrograms through successive measurements.

We establish that each observer represents a distinct subjective knowledge about the universe that evolves not over time, but rather over observations. Thus, a "universe" which fully aligned with the worlds in MWI is composed of an observer "subjective" wave function and the "objective" properties measured from an objective property of other observer. In this context, we unite the concept of the system/observer discrepancy by envisioning a universe composed of observers who measure each other's objective properties. Each observer subjected to measurement has an "objective" wave function dependent on the measuring observer location in the Minkowski-like space. More simply, this "objective" wave function evolves in accordance with the measuring observer's position within the dendrogramic parameter space. Thus, an observer objective property has a different wave function for each location in the dendrogramic parameter space.

Hence, we demonstrate that various interpretations of quantum mechanics emerge when we consider distinct facets of dendrogram evolution. Specifically, these interpretations arise from different viewpoints: one focusing on statistical trajectories across the dendrogram parameter space and the other on a single trajectory.

**1.6 Dendrogram Representation of Events - Preliminaries**

In our pursuit of our objectives, we employ a structured procedure to create a relational framework, specifically a dendrogram. This dendrogram exhibits a unique branching structure through a 2-adic expansion.

Our approach involves representing events, often referred to as Bohr's phenomena, as branches within a dendrogram. A dendrogram, in essence, is a finite tree that serves as an observer's epistemic representation of reality. Figure 1 provides a visual depiction that elucidates the process of constructing such a dendrogramic tree from data. These finite trees are constructed as follows:

1. Data Collection: We commence with the collection of essential data through measurements.

2. Hierarchical Clustering Algorithm: Employing a hierarchical clustering algorithm, along with a selected distance metric and clustering (linkage function) algorithm, we proceed with our methodology.

3. Agglomerative Hierarchical Cluster Tree: Utilizing the chosen distance metric and clustering algorithm, we construct an agglomerative hierarchical cluster tree. Each event, represented by a unique branch, can be encoded as a binary string or a p-adic expansion, based on responses to a series of yes/no questions.

4. Dendrogram Representation: The set of branches within the tree, or the strings that comprehensively describe them, collectively forms the dendrogram. Each branch, associated with a measured event, extends from the root to a leaf, referred to as an edge.

5. Relations Among Events: The existence of a longer common path from the root to the leaf of two branches signifies a closer relation between the corresponding events, determined by the chosen distance metric and clustering algorithm.

The dendrogram representation is facilitated by employing a series of p-adic numbers. Each number encapsulates the relationship between a single event and all other events as perceived by observers. It's worth noting that a dendrogram can only be constructed when there are two or more events, as it fundamentally relies on the relational structure among events. Initially, all N observers possess a basic dendrogram with only two branches.

In the context of an infinite number of events, the ontic description of the event-universe takes the form of an infinite tree. One category of such trees is p-adic trees, characterized by homogeneous branches with $p > 1$ edges extending from each vertex. These trees exhibit an algebraic structure and a corresponding topology consistent with this structure. The p-adic topology adheres to the p-adic ultrametric, satisfying the strong triangle inequality. The p-adic distance between two branches within the tree is determined by their shared root, where a greater common root signifies a shorter distance. As branches represent measurement outcomes of events, whether in a finite or infinite context, the space of events is endowed with a p-adic ultrametric. Consequently, dendrograms encode hierarchical relationships between event measurements based on p-adic distances, rather than the spatiotemporal localization of events.

In this framework, the common root distance serves as the basis for gauging the similarity between measured events. The geometry of this field exhibits intriguing properties, including the characteristic of all triangles being isosceles, a consequence of the strong triangle inequality. Furthermore, by defining "open" and "closed" balls as:

$$B_-(R; a) = \{x : r_p(a, x) < R\}, B(R; a) = \{x : r_p(a, x) \leq R\}$$

Where $r_p(a, x) = |a - x|_p$ is the p-adic distance between the points a and x.

each dendrogram edge can be represented by a 2-adic number:
$edge_i = \sum_{j=0}^{k} a_j \times 2^j$, where $a_j = 0,1$.

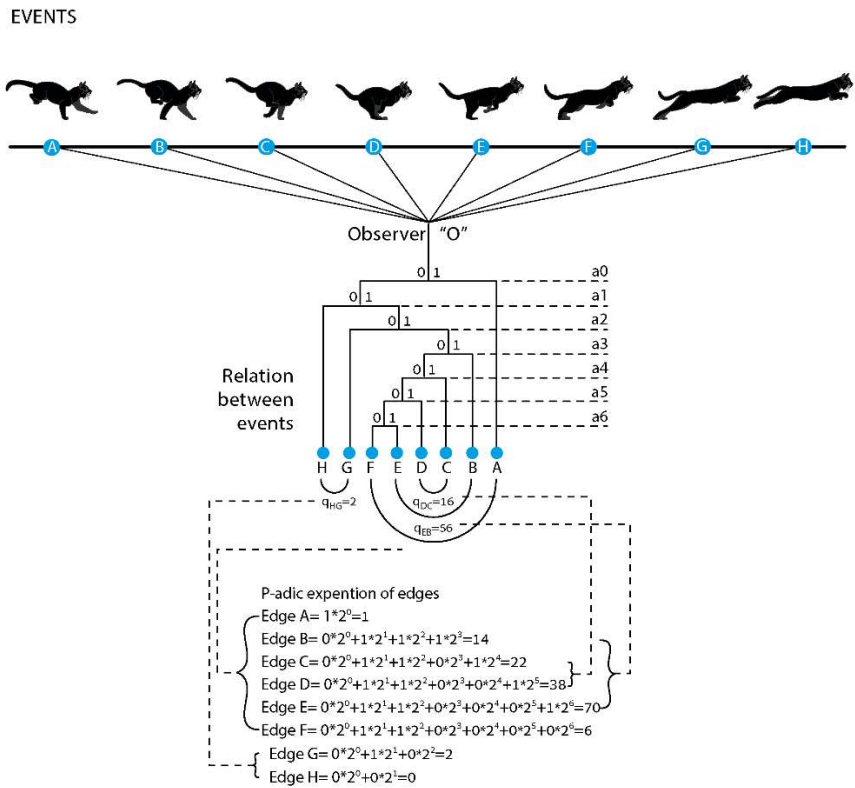

**Figure 1.**
**Relational observation of events.** Observer O discriminates events A–H and constructs an object, a dendrogram, which describes the relations between these events. (**B**) Each edge of the dendrogram is a binary string of 0s and 1s which can be represented as a finite p-adic expansion. Each edge summation of its finite p-adic expansion results in a natural number. Subtracting between two edges' finite p-adic expansion results in "potential gap"—qij.

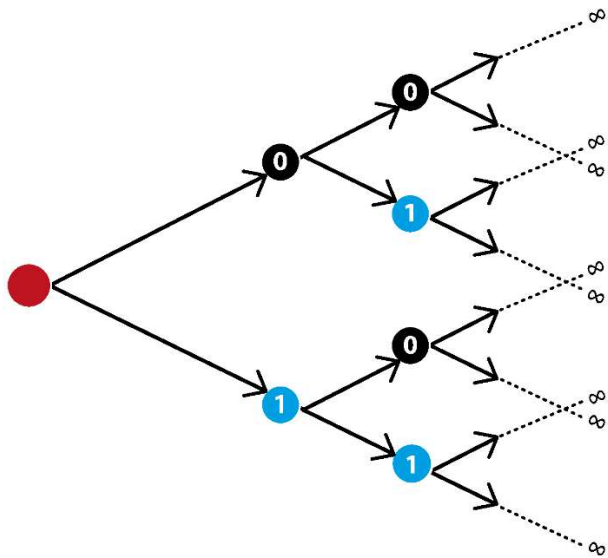

**Figure 2. The 2-adic tree**

### 1.7 Minkowski-like parameter spaces of dendrograms

In the realm of physics and natural sciences, modeling event-observational phenomena has always been a fundamental pursuit. These models often revolve around the concept of events, with space-time acting as a secondary structure for representing these events. DHT has introduced an intriguing shift by proposing a new mathematical framework for understanding the universe's events.

Central to this innovative approach are dendrograms, which as mentioned in the above sections are finite trees that serve as epistemic representations of reality, depicting the events observed by an observer. Unlike traditional models where space-time takes center stage, DHT places events and their relational structures at the forefront. In this context, events are organized into dendrograms, allowing for the exploration of their dynamic interconnections.

In Dendrogramic Holographic Theory (DHT), an innovative concept emerges with Minkowski-like spaces [6], inspired by the Minkowski spacetime of special relativity but tailored for dendrograms and the subjective knowledge of observers. Unlike the determinism of Minkowski spacetime, DHT introduces: *the dendrogramic Minkowski causal structure of observers ensemble relational universes*

For effective modeling in DHT, parameter spaces have been discovered [6] to uniquely define each dendrogram structure, representing an observer's knowledge of the universe. These parameter spaces, remarkably four-dimensional like Minkowski spacetime, offer a versatile tool to accommodate dendrograms of varying sizes within the same framework.

Within these four-dimensional parameter spaces, an information metric has been analytically proven to define what can be described as a dendrogramic "light cone," analogous to the Minkowski spacetime metric, but tailored to characterize to which dendrograms a certain dendrogram structure can evolve to.

By using these four-dimensional parameter spaces we determine the connection between two dendrograms solely on their structures. Dendrograms that demonstrate an observer's transition from one to the other are termed "timelike separated" dendrograms, while those lacking such transitions are labeled "spacelike separated." To be more precise, "timelike/spacelike separated" dendrograms are characterized as follows:

**Definition 1.** *D1* and *D2* are two, **timelike separated**, *different dendrograms with number of events/data collected $e1 \leq e2$ ,respectively, if and only if there exist at least one observer with D1 dendrogram with $e1$ events moves from D1 to D2 upon collecting the next $e2 - e1$ events*

**Definition 2.** *D1 and D2 are two, **spacelike separated**, different dendrograms with number of events/data collected $e1 \leq e2$ ,respectively,*
*if and only if there is no possibility for observers with D1 dendrogram with $e1$ events to move from D1 to D2 upon collecting the next $e2 - e1$ events*

we introduce the concept of a dendrogramic "light cone," by the following definitions:

**Definition 3.** A specific dendrogram D, represented by some 4-dimensional parameter point $\theta$, is identified with numerous observers who share identical relations among all the observations each of them has made of the universe.

Thus a dendrogramic "light cone" is defined as

**Definition 4.** A dendrogramic "future light cone" associated with a particular unique dendrogram encompasses all the potential dendrograms that can emerge from dendrogram D. This holds true irrespective of the observer's identity in relation to D and takes into account all conceivable events and combinations of events they might measure.

Please note that while the causal structure of Minkowski space is not statistical in nature, in DHT, we seek to establish a statistical counterpart referred to as the *"dendrogramic Minkowski causal structure of observers ensemble relational universes "*.

Where a *universe is defined as*

**Definition 5.** A universe is an observer, current, epistemic, relational knowledge (information) he acquired on the ontic universe by measuring some finite amount of events

Crucially, the parameter spaces discovered within DHT are inherently subjective. They primarily represent the subjective and observer-dependent knowledge of an individual observer about the universe. This subjectivity is not a limitation but a fundamental aspect of DHT, distinguishing it from classical spacetime physics and highlighting its capacity to embrace the diverse perspectives of different observers.

## 2. Model and Results.

We already proved there exist a Minkowski-like parameter space of dendrograms $\boldsymbol{\theta}$ where each coordinate in $\boldsymbol{\theta}$ has a light cone meaning an observer subjective knowledge of the universe which is represented by $\boldsymbol{\theta}$ can only reach a sub-section of the parameter space by measuring events [6]. Each $\boldsymbol{\theta}$ uniquely defines a unique subjective dendrogram structure of an observer. The same structure, or same $\boldsymbol{\theta}$, is the current subjective dendrogram of a fraction of N observers in the universe (where N→∞) upon observing M events (generally these M events are different).

An observer is uniquely defined by its world line curve, whether it involves acceleration or not, within the Minkowski-like parameter space, we inherently acknowledge that each observer will eventually perceive a distinct ontic relational perspective of the universe. It's essential to note that the presence of two observers sharing the same world line is not permissible, as this would essentially equate them to the same single observer, in accordance with the principles of Leibniz.

An observer ontic view of the universe is dependent on which measurements he will preform. Thus his world line through the subjective parameter space is defined by its measurements and vice versa.

In summary, following the Leibnitz principle, the uniqueness of an observer's perspective on the universe is synonymous with the idea that each observer will conduct a distinct and infinite set of measurements, and this, in turn, equates to the notion that every observer possesses a unique world line within the parameter space.

Hence, when we consider an infinite number of observers (N→∞), each of them is uniquely defined. We can create an infinite p-adic tree by posing an unlimited series of yes/no questions, with each branch representing an objective property of an observer. These questions may pertain to distinctions in world lines or variations in sets of measurements, which are synonymous. Thus a p-adic number is an objective ontic property of an observer.

The events an observer measures are other observers ontic objective property. These observers are contained in its past light-cone (meaning if we go back to the two edges dendrogram $\boldsymbol{\theta}$ coordinate it represents all N observers). Thus the ontic objective properties of the observers are accessible to any observer as they are the p-adic number field. In this way we connect the ontic (the p-adic infinite tree which may be considered situated in plato world of ideas) with the epistemic subjective view of an observer.

More over in this model the observer/system discrepancy unites into an all observer universe. we might consider an observer, in this model, as every physical entity which interacts with another an atom, electron etch..

## 2.1 The observer subjective wave function

We briefly outline the construction of the observer subjective wave function ( which was studied in our recent work [5] ):

We start with constructions of the "differences pdf", $\rho$. This will be our fundamental distribution. Thus, given a dendrogram that describes, p-adicaly, the relations between m events we calculate all possible pairwise differences of the p-adic edges representation and represent them as events through monna map as above:

$$q_{ik} = (event_i - event_k) \qquad (1)$$

Where for p = 2,

$$edge_i = \sum_{j=0}^{k} a_j \times 2^j \rightarrow event_i = \sum_{j=0}^{k} a_j \times 2^{-j-1}, \ a_j = 1,0, \qquad (2)$$

the Monna map maps natural numbers into rational numbers belonging the segment [0,1]. A dendrogram is mapped into a subset of [0,1]. This map can be extended to the infinite p-adic tree where its branches are represented by infinite series; for p=2,

$$edge_i = \sum_{j=0}^{\infty} a_j \times 2^j \rightarrow event_i = \sum_{j=0}^{\infty} a_j \times 2^{-j-1}, \ a_j = 1,0, \qquad \text{as in (2)}$$

The latter is important for considering the subjective knowledge of an observer in the limit infinitely many events collected by observations; they are portrayed on an infinite 2-adic tree. Thus, from all $q_{ik}$ we have a discrete "difference pdf", $\rho$, which is the fraction of each unique value of $q_{ik}$ defined For the set of the unique $q_{ik}$ values Q we have:

$$\rho_j = \frac{(number\ of\ different\ q_{ik}\ that\ equal\ Q_j)}{total\ number\ of\ q_{ik}} \qquad (3)$$

We then define the differences energy, which will be equivalent to the usual kinetic energy, we define the equation:

$p = \frac{1}{N}\sum_{k \neq j} q_{jk}$ in which N is the number of possible $q_{jk}$. \qquad (4)

This formula represents the mean difference between all pairwise events. Thus, $p$ takes the part as the momentum in Bohmian mechanics.

Now, we introduce the phase $S$ as the new parameter of the model:

We set $p = \partial S$ and $\partial S = e^{iS}$. (5)

We can then calculate the "differences energy" as the real part of:

$$T_{differences\ energy} = \frac{1}{N}\Sigma_{j \neq k}(q_{jk})^2 \qquad (6)$$

Where $T_{differences\ energy}$ serves as our equivalent of kinetical energy

The continuum approximation of $T_{differences\ energy}$ will be obtained by noticing

$$p_j = \rho(Q_j)(Q_j)^2 = \partial S_j$$

(7)

Thus:

$$T_{differences\ energy} = \int (\rho(Q_j)(Q_j)^2)\rho(Q_j)dQ_j = \int (\partial S_j)^2 \rho(Q_j)dQ_j$$

(8)

By using the proceedures outlined in [5,36] and[36] we derive the action principle:

$$A(S,\rho) = \int d(Dendrogram)\{\int \frac{d(S)}{dDendrogram}\rho(Q)dQ + \int (\partial S)^2 \rho(Q)dQ - v(Q) + U(Q)\}$$

Where $v(Q) = Z_V \int \rho(Q)(\frac{(\partial \rho(z))^2}{(\rho(Q))^2})\,dQ$ and $U(Q) = \int \rho(Q)dQ$

(9)

Where $U(Q) = \int \rho(Q)dQ$

The Hamilton Jacobi equation of the action (notice we are on the straight line [ 0 1] now in all our pdfs ).

$$-\dot{S} = (\partial_{edge}S)^2 + U + U^Q \qquad (10)$$

In which $U^Q = (\Delta^2\sqrt{\rho})/\sqrt{\rho}$ is the quantum potential

And $U$ is the potential

And $\dot{S} = S(present\ dendrogram) - S(previous\ dendrogram)$

All parameters are well-defined.

The probability conservation law is

$$\dot{\rho} = \partial_{edge}\,(\rho\partial_{edge}S)^2$$

In which $\dot{\rho} = \rho(present\ dendrogram) - \rho(previous\ dendrogram)$ (11)

Equations 10 and 11 are the real and imaginary parts of the Schrodinger equation for $\psi_{subjective} = \sqrt{\rho}e^{iS}$.

We emphasize that the subjective wave function, $\psi_{subjective}$, is completely dependent on the measurements the observer is preforming thus for a set $M=\{m_1, m_2.. m_i\}$, $\psi_{subjective}(M) = \psi_{subjective}(\boldsymbol{\theta})$ where different sets M can have same $\boldsymbol{\theta}$.

an observer may be characterized uniquely by its infinite set of measurements he preforms. In that sense these infinite set of measurements are its world line curve (accelerated or not) on the background of spacetime ($\boldsymbol{\theta}$), each observer will ultimately obtain different ontic relational view of the universe (two observers with same world line are not allowed as they are the same observer if we postulate Leibnitz principle). This view is dependent on the information about events, localized in the background of spacetime, transmitted to the observer moving with acceleration or without it in spacetime. The transition of this kind of measurements "world line" into the dynamical evolvement of the subjective wave function, $\psi_{subjective}(\boldsymbol{\theta})$, is trivial.

## 2.2 An observer objective property transformation to wave function

Let us consider the situation where all observers at a particular $\boldsymbol{\theta}$ measure a particular observer $O_{B_k}$. The objective property of $O_{B_k}$ is composed as follows : the sum of the finite p-adic expansion with $x_0\ x_1\ x_2...\ x_k$, is $Z_{B_k}$. This p-adic expansion can be transformed by monna map to a rational number $q_{B_k} \in [0\ 1] \subset \mathbb{Q}$ in the

Each of the observers at $\boldsymbol{\theta}$ measure the same $q_{B_k}$ and incorporate it subjectively into their dendrogram. Thus each observer at $\boldsymbol{\theta}$ will incorporate it as a finite branch with different p-adic expansion that by the monna map will be identified with a rational number value, we call event, on the interval $[0\ 1]$. So for $\boldsymbol{\theta}$ $O_{B_k}$ is a distribution of possible rational numbers. we thus identify $O_{B_k}$ as a distribution $\rho_{B_k}(x, \boldsymbol{\theta})$ on the interval $x \in [0\ 1]$. From these possible events values we can constract the objective wave function of $O_{B_k}$ for $\boldsymbol{\theta}$ . the procedure to construct the objective wave function of $O_{B_k}$ follows the same lines as in section 2.1 (equations 1-10) but now in equation 2, $event_i = \sum_{j=0}^{\infty} a_j \times 2^{-j-1}$, $a_j = 1,0$ will be the event of measuring observer $O_{B_k}$ objective property (which is the ball value that represents $O_{B_k}$) by some other observer at $\boldsymbol{\theta}$.

$\psi^{O_{B_k}}(\boldsymbol{\theta}) = \sqrt{\rho_{B_k}(x,\boldsymbol{\theta})}e^{iS(\boldsymbol{\theta})}$. So $\psi^{O_{B_k}}(\boldsymbol{\theta})$ is a relational wave function (in relation to a certain $\boldsymbol{\theta}$). For each different $\boldsymbol{\theta}$ it dynamically changes. this wave function of $O_{B_k}$, is the transformation of an objective property to an ensemble of subjective properties (dependent on the ensemble of observers measuring it) in relation to $\boldsymbol{\theta}$. Where $\boldsymbol{\theta}$ is inherently subjective, e.g. it represents each of the observers at $\boldsymbol{\theta}$ with same subjective knowledge about the universe.

## 2.3 measurement

So we have an observer at $\boldsymbol{\theta}$ that has a subjective dendrogram which again (as in our previous study and section 2.1) we can construct from it a subjective wave function $\psi^{\boldsymbol{\theta}}$ for the $\boldsymbol{\theta}$ coordinate. The fraction $b_j$ of observers at $\boldsymbol{\theta}$, all with same dendrogram, that will measure $O_{B_k}$ and will transform upon this measurement to $\boldsymbol{\theta_1}$ thus they will be before measurement

$$\psi^{\theta^{b_j}+O_{B_k}} = \sum_i a_i \phi_i^{O_{B_k}} \psi^{\theta}$$

And after

$$\psi^{\theta_1+O_{B_k}} = \sum_i a_i \phi_i^{O_{B_k}} \psi^{\theta_1}$$

So the state of $\boldsymbol{\theta}(M)$ (M is the number of edges of the dendrogram that encodes the coordinate $\theta$) at $\boldsymbol{\theta}(M+1)$ upon all observers in $\boldsymbol{\theta}$ measuring $O_{B_k}$ is:

$$\psi^{\theta(M+1)+O_{B_k}} = \sum_j \sum_i a_i \phi_i^{O_{B_k}} b_j \psi^{\theta_j}$$

Where $\boldsymbol{\theta_j} \neq \boldsymbol{\theta}$ and $j$ runs from 1 to u

We can now generalize the situation into $\boldsymbol{\theta}(M)$ measuring several $O_{B_k}$'s so $k = \{1,2..h\}$
The combined distribution of $O_{B_{1,2..h}}$ with respect to $\boldsymbol{\theta}$ is $\rho_{B_{1,2..h}}(x,\boldsymbol{\theta})$ and

$\psi^{O_{B_{1,2..h}}}(\boldsymbol{\theta}) = \sqrt{\rho_{B_{1,2..h}}(x,\boldsymbol{\theta})}e^{iS(\boldsymbol{\theta})}$ with eigen values $\phi_i^{O_{B_{1,2..h}}}$ inserting it to the equation above we have:

$$\psi^{\theta(M+1)+O_{B_{1,2..h}}} = \sum_j \sum_i a_i \phi_i^{O_{B_{1,2..h}}} b_j \psi^{\theta_j}$$

We can even generalize to a region of the parameter space

So let $\bar{\theta}(G) = \{\theta_1(k), \theta_2(l), \theta_3(f)....\}$, $G = k, l, f ...$

And the combined distribution of $O_{B_{1,2..h}}$ with respect to $\bar{\theta}$ is $\rho_{B_{1,2..h}}(x, \bar{\theta})$ and

$\psi^{O_{B_{1,2..h}}}(\bar{\theta}) = \sqrt{\rho_{B_{1,2..h}}(x, \bar{\theta})} e^{iS(\bar{\theta})}$ with eigen values $\phi_i^{O_{B_{1,2..h}}}(\bar{\theta})$

Then we have:

$$\psi^{\bar{\theta}(G+1)+O_{B_{1,2..h}}(\bar{\theta})} = \sum_j \sum_i a_i \phi_i^{O_{B_{1,2..h}}} b_j \psi^{\bar{\theta}_j}$$

We then can have by Everett second rule another measurement of different set of observers $O_{B_L}$ where $L \neq \{1,2 ... h\}$ resulting in

$$\psi^{\bar{\theta}(G+2)+O_{B_{1,2..h}}(\bar{\theta})} = \sum_j \sum_l \sum_i c_l \varphi_l^{O_L} a_i \phi_i^{O_{B_{1,2..h}}} b_j \psi^{\bar{\theta}_j}$$

We then need to consider the situation where some observer at $\bar{\theta}$ will measure an observer $O_{B_m}$ that he already measured so his dendrogram will not change and so does his wave function. thus

$$\psi^{\bar{\theta}(G+1)+O_{B_{1,2..h}}(\bar{\theta})+\bar{\theta}} = \sum_j \sum_i a_i \phi_i^{O_{B_{1,2..h}}} b_j \psi^{\bar{\theta}_j}$$

Where now $\bar{\theta}_j$ $runs$ $from$ $1$ $to$ $u + size$ $of$ $\bar{\theta}(G)$

We now compare with Everett's interpretation

We note that for $\bar{\theta}$ there is no longer any independent state of the observers $O_{B_{1,2..h}}(\bar{\theta})$ or the $\bar{\theta}(G+1) + \bar{\theta}$. However each element of the superposition, $\phi_i^{O_{B_{1,2..h}}} \psi^{\bar{\theta}_j}$, is in a particular eigenstate of $\bar{\theta}$, $\psi^{\bar{\theta}_j}$, and furthermore the $\bar{\theta}$-$O_{B_{1,2..h}}(\bar{\theta})$ state, $\phi_i^{O_{B_{1,2..h}}} \psi^{\bar{\theta}_j}$, describes all observers at $\bar{\theta}_j$ as definitely perceiving that particular system state.

Please compare with Everett's thesis:

> We note that there is no longer any independent system state or observer state, although the two have become correlated in a one-one manner. However, in each *element* of the superposition (2.3), $\phi_i \psi^O_{i[...,a_i]}$, the object-system state is a particular eigenstate of the observer, and *furthermore the observer-system state describes the observer as definitely perceiving that particular system state.*[1] It is this correlation which allows one to maintain the interpretation that a measurement has been performed.

We note that for each observer at the superposition combination $\phi_i^{O_{B_{1,2..h}}} \psi^{\bar{\theta}_j}$ the encoded eigenvalue $\alpha_i$ i of $\phi_i^{O_{B_{1,2..h}}}$ is encoded in $\psi^{\bar{\theta}_j}$ subjectively the same as in the $\psi^O_{i[...,a_i]}$ of Everett's.

In contrast to Everett the memory $[....\alpha_i]$ is not constant but subjectively changes so we should note it as in the next measurements as $[....\alpha_i''..]$.

In this formalism $\phi_i^{O_{B_{1,2..h}}}$ Is the objective property transformation to the subjective measurement thus while $\psi^{\bar{\theta}_j}$ is the purely subjective knowledge of an observer of the universe. each world, in the MWI meaning, is a world line of objective observetions in superposition with an observer subjective wave function. We can identify the world line as the $\phi_i^{O_{B_{1,2..h}}}$, $\varphi_i^{O_{B_{1,2..h}}}, \eta_i^{O_{B_{1,2..h}}}$ .... sequences.

Lets define the equivalent of the MWI relative state taken from Everett's thesis:

> a composite system $S = S_1 + S_2$ in the state $\psi^S$. To every state $\eta$ of $S_2$ we associate a state of $S_1$, $\psi^\eta_{rel}$, called the relative state in $S_1$ for $\eta$ in $S_2$, through:
>
> (1.9)   DEFINITION. $\psi^\eta_{rel} = N \sum_i (\phi_i \eta, \psi^S) \phi_i$ ,

So we have

For the one $O_{B_k}$ and single $\psi^{\theta_j}$ we decompose $\psi^{\theta_j}$ into it's eigen function $\varphi_k$

$$\psi^{\theta_j(M+1)+O_{B_k}} = \sum_i a_i \phi_i^{O_{B_k}} \psi^{\theta_j} = \sum_k \sum_i a_i \phi_i^{O_{B_k}} c_k \varphi_k$$

$$\phi_i^{O_{B_k}} \varphi_k$$

$$\psi^{\theta(M+1)+O_{B_k}} = \sum_j \sum_i a_i \phi_i^{O_{B_k}} b_j \psi^{\theta_j} = \sum_j \sum_k \sum_i a_i \phi_i^{O_{B_k}} b_j c_{k_j} \varphi_{k_j}$$

So the relative state in $O_{B_k}$ for $\theta_j$ is

$$\psi_{rel}^{\theta_j} = \frac{1}{Z} \sum_i \sum_k (\phi_i^{O_{B_k}} c_{k_j} \varphi_{k_j}, \psi^{\theta(M+1)+O_{B_k}}) \phi_i^{O_{B_k}}$$

Where Z is a normalization constant. Thus $\psi_{rel}^{\theta_j}$ correctly gives the conditional expectation of all operators in $\psi^{O_{B_k}}$ conditioned by the state $\psi^{\theta_j}$ in $\psi^{\theta(M+1)}$

the relative state in $\theta$ for $\phi_i^{O_{B_k}}$ is

$$\psi_{rel}^{\phi_i^{O_{B_k}}} = \frac{1}{Z} \sum_j \sum_k (\phi_i^{O_{B_k}} c_{k_j} \varphi_{k_j}, \psi^{\theta(M+1)+O_{B_k}}) c_{k_j} \varphi_{k_j}$$

Where Z is a normalization constant. Thus $\psi_{rel}^{\phi_i^{O_{B_k}}}$ correctly gives the conditional expectation of all operators in $\psi^{\theta(M+1)}$ conditioned by the state $\phi_i^{O_{B_k}}$ in $O_{B_k}$.

## 3. Discussion

In this study we showed the natural unification of the single observer "subjective" Bohmian mechanics which emerged under the framework of DHT with the MWI formalism. This unification was achieved in the emergent Minkowski-like causal structures, an emergent consequence of ensemble of observers. These Minkowski-like causal structures are inherently subjective in nature, meaning they characterize the observer subjective relational knowledge of the universe.

The model applied in this study unites the observer/system discrepancy into ``observers universe''. Measurements are performed by observers on observers. Such ``observers universe'' is dynamically changing and is background independent since the space itself is emergent.

Another intriguing consequence of our model is the emergent transformation of any objective property into a subjective knowledge of an observer. Thus, the transformation of an objective property into a wave function immediately indicates that the wave-function is a representation of ensemble of subjective knowledge of the observers. This is in contradiction to Bohr's view, where the wave function is objective and gaining information about it by measurement is always subjective.

Moreover, the emergent consequences of DHT are in alignment with Rovelli's RQM postulates (explained in the introduction). We emphasize that DHT does not need Rovelli's postulates, rather they emerge out of Leibnitz principle. This connection should be further developed analytically.

It is encouraging that two interpretations of quantum theory are simply emergent under the framework of DHT. Alongside the alignments with RQM postulates, this study might suggest a unification of all Quantum theory interpretation under the DHT framework. More research needs to be conducted on other interpretations and their connections to DHT.

One more consequence of this study is the close linkage between the world line of an observer to the measurements he is preforming. Thus, dynamics of the wave function is not a function of space and time but a function of measurements which in turn can define world line through space time in the classical studies of DHT

In this model, it's important to highlight that the ontic objective properties of observers are within reach for any observer, as they form part of the p-adic number field. This connection effectively bridges the ontic realm, represented by the p-adic infinite tree, which could be conceptualized as residing in Plato's world of ideas, with the epistemic subjective perspective of each individual observer.

The objective, complete, ontic world line, along with its p-adic representation attributed to each observer, remains entirely static and unchanging in nature, devoid of any dynamics. Surprisingly, out of subjectiveness considerations, namely on the subjective 4-dimensional Minkowski-like parameter space, we wittnese the emergence of dynamics. This dynamics of the objective static property is dependent on the subjective parameter space $\theta$, in the form of a wave function $\psi^{O_{B_k}}(\theta)$ described in secion 2.2. Thus subjectivity in our model is the source of all apparent dynamics.

The overall analytical and numerical studies of DHT suggest some unification between quantum and classical paradigms of reality. More studies are needed to emerge gravitation and the usual entities like fields and particles of usual physics.

**Funding**

This research did not receive any specific grant from funding agencies in the public, commercial, or not-for-profit sectors.

**Availability of Data and Materials**

Data sharing is not applicable to this article as no new data were created or analyzed in this study.

**Conflicts of Interest**

None of the authors has any conflict of interest to disclose.

**Acknowledgements:** Katy and Salomon Pollar-Luchevsky for intelligent discussion and support, and Carlos Baladron, Brank Dragovich, and Igor Volovich for kind advices on the improvement fo presentation.